\documentclass[aps,twocolumn,prd,superscriptaddress,preprintnumbers,showpacs]{revtex4}
\usepackage{graphicx}
\usepackage{bm}
\usepackage{amsmath}
\usepackage{amssymb}
\usepackage{amsthm}
\newcommand{\be}{\begin{equation}}
\newcommand{\ee}{\end{equation}}
\newcommand{\bea}{\begin{eqnarray}}
\newcommand{\eea}{\end{eqnarray}}
\newcommand{\mpl}{M_{\rm pl}}

\begin{document}

\title{Diffusing non-local inflation:\\
Solving the field equations as an initial value problem}

\author{D. J. Mulryne}
\email[]{D.Mulryne@damtp.cam.ac.uk}
\affiliation{Department of Applied Mathematics and Theoretical Physics, Wilberforce Road, Cambridge, CB3 9AN, UK}
\author{N. J. Nunes}
\email[]{nunes@damtp.cam.ac.uk}
\affiliation{Department of Applied Mathematics and Theoretical Physics, Wilberforce Road, Cambridge, CB3 9AN, UK}

\date{\today}

\begin{abstract}
There has been considerable recent interest in solving
non-local equations of motion which contain an infinite number of
derivatives.
Here, focusing on inflation, we review how 
the problem can be
reformulated as the question of finding solutions to a diffusion-like partial differential equation with
non-linear boundary conditions. Moreover, we show that this diffusion-like equation,
and hence the non-local equations, can be
solved as
an initial value problem once non-trivial initial data consistent with
the boundary conditions is found. This is done  
by considering linearised  equations about any field value, for which we 
show that obtaining solutions using the diffusion-like 
equation is equivalent to
solving a local but infinite field cosmology. These local fields 
are shown to consist of 
at most two canonically normalized or phantom fields together with an
infinite number of quintoms. We 
then numerically solve the diffusion-like equation for the full
non-linear case for two string field theory motivated models.
\end{abstract}

\maketitle

\section{Introduction}
\label{introduction}

Non-local field equations which contain infinitely many derivatives
have recently received considerable attention in the context of
cosmology \cite{biswas, NonLocalCosmolOthers, darkTachyon, koshelev, calcagni1,calcagni2, lidsey, non-Gaussian, ludaLinear, ludaBoundary}. 
In particular, the effective actions of a number of models derived
from string field theories have been studied.  Examples are 
the ground state tachyon in 
bosonic cubic string field theory
(BCSFT) \cite{sft} and cubic super-string field
theory (CSSFT) (see \cite{reviews} for reviews), 
and in $p$-adic string theory \cite{p-adic}. 

Keeping these particular examples in mind, we will for the majority of 
this paper keep the model we consider as general
as possible so that our methodology can be applied widely.  Moreover,
though we study only single field models it is clear that our method
can easily be extended to encompass additional fields.  As we will discuss later, the method can also be generalized to solve cosmological perturbations in 
non-local models, and this is a major motivation for its development. 
The key
ingredient we require is that the non-local operator which the field
equations contain be of the form $\exp(-\alpha \Box)$ where $\alpha$
is a constant and $\Box$ the d'Alembertian operator. As we will show, 
an operator of
this form implies that
finding a solution to the non-local field equations is
equivalent to 
finding a solution to a partial differential equation (pde) which
resembles the diffusion equation. This idea is not new (see for example related work in Refs. \cite{vladimirov, calcagni1,calcagni2}), 
and has recently been employed in papers very closely connected to this work 
\cite{ludaBoundary}. As we discuss in more
detail later,  this
recent work solves the diffusion-like pde as a boundary value problem, that is, it specifies the initial and final state of the field and seeks solutions which interpolate between these states.  Our interest, however, is in how the field equations, and hence
the diffusion-like pde, 
can be solved as an initial value problem, specifying 
only initial conditions for the field's evolution.

For linear models an initial value formulation does not require us to 
consider the diffusion-like equation directly,  
since, as we will review, a correspondence between 
the non-local field equations with one field and local field equations
with an infinite number of fields can be made. Thus the problem of 
specifying initial
data for the solution of the non-local field equations is reduced to
specifying initial conditions for these local fields \cite{ludaLinear, koshelev} (also see \cite{ivp}).  
For an isotropic
cosmology these initial conditions correspond to the initial value and velocity for each
field (we should note that not all fields need to play an active role --
some can simply be set to zero). In our work we extend the
understanding of this correspondence with an infinite number of local fields by 
generalizing the linear
models previously considered with quadratic terms, to include potentials with linear terms as well. Moreover, we show that if we require real solutions,
the fields can be interpreted as containing at most two canonically
normalized fields or phantom fields, together with an infinite number of
quintoms.  
Let us review the definition of these fields. 
Here and throughout we use the signature of the metric $(-,+,+,+)$.
A field $\varphi$ coupled to gravity with the action
\be
\label{RealAct}
S=\int d^4x \sqrt{-g} \left [ \frac{\mpl^2}{2} R +  \left
  (s\frac{1}{2} \varphi \Box \varphi - V(\varphi) \right ) \right ] \,
\ee
represents a canonically normalized field when $s=1$, but a
phantom when $s=-1$.
A quintom is defined as a combination of one canonically normalized field and one phantom field interacting through a shared potential and has the
action 
\be
S=\int d^4x \sqrt{-g} \left [ \frac{\mpl^2}{2} R +  \left
  (\frac{1}{2} \chi \Box \chi  - \frac{1}{2} \sigma \Box \sigma - V(\chi, \sigma) \right ) \right ] \,.
\ee
Phantom fields violate the weak energy condition (the name together with a major industry began with \cite{Caldwell:1999ew}), and hence their energy
density can increase with time.  
Quintoms, on the other hand, have the property that
their energy density oscillates with time. 
Pairs like these where put forward as a simple way to explain the apparent transition of the equation of state of the universe from $\rho+P > 0$ at large redshifts to $\rho+P<0$ at small redshift and were coined as ``quintoms" \cite{quintoms}.  Phantom fields raise concerns because 
naively they are unstable to vacuum decay (see for example \cite{Carroll:2003st} for constraints), though the 
question here is complicated as they enter at an effective level. For the purposes 
of this paper, which is to determine how to solve the equations of motion, 
we will leave the issue of physicality 
of phantoms aside. One should also be aware of the possibility of classical instabilities  (see Ref.~\cite{ostro}, though the construction in this work requires a finite number of derivatives in the theory).

The major step forward made in our work is to show that, for linear models, solving the
non-local equations using the 
correspondence technique is completely equivalent to solving the 
equations using
the diffusion-like pde.  Furthermore, we determine what initial data 
for the pde is
equivalent to introducing each of the local fields.  By linearising
a non-linear model about a particular field value, we will be able to
use these results to specify initial data for
non-linear models, which we then evolve into the full non-linear
regime using the diffusion-like equation.  Moreover, the understanding we
gain from the linear models allows us to give an interpretation of 
 the initial conditions for the pde, even in the non-linear case, 
in terms of which fields are present initially.

Our paper is structured as follows: In section II we introduce the non-local model we will be 
considering.  Section III reviews the
difficultly in solving non-local field equations as an initial value
problem before introducing the diffusion-like pde. Section IV considers linear models and their correspondence with an infinite 
field local cosmology, and section V and VI shows how this understanding leads to suitable initial data for solving 
the diffusion-like equation. Numerical solutions are presented in section VII for two examples. In section VIII we discuss 
how our method can be extended to solve perturbation equations in non-local cosmology, and finally in section IX 
we conclude and outline future directions. 

\section{Non-local Lagrangians}
We begin by considering a non-local action for a single 
scalar field $\psi$ minimally coupled to gravity, of the form:
\be
\label{Lag}
S=\int d^4x \sqrt{-g} \left [ \frac{\mpl^2}{2} R + \gamma^4 \lambda \left
  (\frac{1}{2} \psi F(\Box) \psi - V(\psi) \right ) \right ] \,,
\ee
where $\Box$ represents the d'Alembertian operator
\be
  \Box \equiv \frac{1}{\sqrt{-g}} \partial_\mu \left( \sqrt{-g} \, g^{\mu
  \nu}\partial_\nu \right) \,,
\ee  
and in this work we will take 
\be
F(\Box)= -(1+4\xi^2\alpha \Box) e^{-\alpha \Box} \,,
\ee
which covers a wide range of theories appearing in the literature
including the $p$-adic string,  the
bosonic string field theory (BCSFT) and the  
cubic superstring field theory (CSSFT). $\xi^2$, $\gamma^4$ and $\alpha$ are constants, fixed by the particular theory of interest, later we will see a number of concrete examples but for now we can treat them as free parameters. We should note that in the string field theory context our coupling to gravity has been introduced essentially by hand (as it has been in all other works on non-local cosmology) as the minimal covariant generalisation of the flat space Lagrangian.

The scalar field equation which follows from the action (\ref{Lag}) is 
\be
\label{ScalarField}
(1+4\xi^2\alpha \Box) e^{-\alpha \Box} \psi= -\frac{d V(\psi)}{d \psi}\,~,
\ee
while the Einstein field equations take the usual form 
\be
\label{Einstein}
G_{\mu \nu}= \frac{8  \pi}{\mpl^2} T_{\mu \nu} \,,
\ee
with $G_{\mu \nu}$ being the Einstein tensor, 
and the non-local stress energy tensor( first derived in Ref. \cite{se}) is given by 
\begin{eqnarray}
\label{StressEnergy}
&~& \frac{T_{\mu \nu}}{\gamma^4\lambda} = - \frac{2}{\gamma^4\lambda \sqrt{-g}}\frac{\delta S_\psi}{\delta g^{\mu\nu}} =  \nonumber \\ 
&~&
\frac{1}{2} g_{\mu \nu} \left[ \psi e^{-\alpha\Box}\psi + 2 V(\psi) -4\xi^2\alpha \partial_\sigma\psi \partial^{\sigma}\left(e^{-\alpha \Box}\psi\right) +  
\right. \nonumber \\  
&~&
 \alpha \int_0^1 d\tau \left( e^{-\tau\alpha\Box}
    \left(1+4\xi^2\alpha\Box\right)\psi \right) \left(\Box
e^{-(1-\tau)\alpha\Box}\psi\right)+ \nonumber \\
&~&
\left. \alpha \int_0^1d\tau \partial_\sigma \left(e^{-\tau\alpha
    \Box}(1+4\xi^2\alpha\Box)\psi\right) \partial^\sigma 
\left(e^{-(1-\tau)\alpha\Box}\psi\right) \right] - \nonumber \\ 
&~&
\alpha  \int_0^1 d \tau \partial_\mu \left(e^{-\tau\alpha
    \Box}(1+4\xi^2\alpha\Box)\psi \right) \partial_\nu 
\left(e^{-(1-\tau)\alpha\Box}\psi\right) +  \nonumber \\
&~&  4\xi^2\alpha\partial_\mu \psi\partial_\nu
    \left(e^{-\tau\alpha\Box}\psi \right) \,. 
\end{eqnarray}

\section{Solving as an initial value problem}
\label{cauchyproblem}
In order to solve a scalar field equation with a large but finite number of derivatives, for example $n$ derivatives of $\psi$, we would need to specify an initial value for $\psi$, $\dot{\psi}$, ..., $d^{n-1} \psi/dt^{n-1}$, which can be given independently of one another.
For a system with an infinite number of derivatives, however, this would require an infinite number of initial
conditions and moreover it would imply that in principle any solution is possible. This argument follows since freedom to specify 
each of the infinite number of initial values 
is equivalent to the freedom in specifying a full Taylor series solution 
for the field's evolution (assuming of course that the solution is analytic in the neighbourhood of the initial conditions). 
This is not the case, however, since the scalar field equation 
itself imposes an infinite number of constraints  
on the allowed initial values of the derivatives (as is discussed at length in Ref.~\cite{moeller}). For example, for a
model 
with $V(\psi) = -\psi^4/4$ and 
$\xi^2 =0$, let us take some initial value 
$\psi_0=\psi(t=0)$ and fix the initial conditions such that all
derivatives are zero except 
$\Box\psi|_{t=0}$.  
Equation  (\ref{ScalarField}) then implies the condition $-\alpha \Box\psi|_0 = \psi_0^3-\psi_0$, however,
it can be verified, in Minkowski space, that 
operating twice on both sides of Eq.~(\ref{ScalarField}) by
the d'Alembertian operator implies that either 
$\Box\psi|_0=0$ or $\psi_0 = 0$ 
which is 
incompatible with the specified initial conditions.  
This simple example illustrates that differentiating the equation of motion imposes additional constraints upon the initial data. In general, 
repeatedly applying the d'Alembertian operator to both sides 
of the scalar field equation 
generates an infinite number of constraints on the initial data. Our
initial data cannot therefore be arbitrary, and attempting to specify
initial data which satisfies all the constraints seems hopeless.  
This is an important point to which we will refer later.

\subsection{The diffusion-like equation}
It seems that the naive formulation of the initial value problem
for non-local field equations is intractable.  
We will now show that we can reformulate the problem in an alternative way by recalling a connection between the non-local field equations and a diffusion-like pde \cite{vladimirov,calcagni1,calcagni2,ludaBoundary}.
Although our discussion can be easily extended to an arbitrary form of
the metric, we will at this point specialize to the homogeneous and isotropic 
Friedmann-Robertson-Walker metric, as our primary interest is non-local cosmology. In this case, $\psi$ is a function only of time, and the d'Alembertian operator becomes 
\be
\Box= -\frac{d^2}{dt^2}-3 H(t) \frac{d}{dt}\,~,
\ee
where $H$ is the Hubble rate, $H \equiv \dot{a}/a$, given by the usual Friedmann equation
\be
\label{Fried}
H^2=\frac{8 \pi }{3 \mpl^2}\rho \,,
\ee
where for our non-local theory, $\rho$ is given by the stress energy tensor (\ref{StressEnergy}) such that $\rho = -T_{00}$, and takes the form
\begin{eqnarray}
\rho  &=& \frac{\gamma^4 \lambda}{2} \left[  \psi e^{-\alpha \Box}
\psi + 2V(\psi)  - 4\xi^2 \alpha  \dot{\psi} \, \partial_t \left ( e^{-\alpha \Box} \psi
  \right ) \right. + \nonumber \\
 &~&\alpha \int_0^1 d\tau \, \left( e^{-\tau\alpha\Box}(1+4\xi^2 \alpha  \Box) \psi \right) \left(\Box e^{-(1-\tau)\alpha \Box} \psi \right) +  \nonumber\\
 &~& \left. \alpha \int^1_0 d\tau \, 
\partial_t \left(e^{-\tau\alpha\Box}\left(1+4\xi^2 \alpha \Box\right) \psi \right)
\partial_t \left(e^{-(1-\tau)\alpha\Box}\psi \right) \right] \,, 
\nonumber \\
\end{eqnarray}
and the Raychaudhuri equation also takes the usual form
\be
\label{Hdot}
\dot{H}= - \frac{4 \pi}{\mpl^2} \left(\rho + P \right) \,,
\ee
where the pressure $P$ is given by
\begin{eqnarray}
P  &=& \frac{\gamma^4 \lambda}{2} \left[  -\psi e^{-\alpha \Box}
\psi - 2V(\psi)   - 4\xi^2 \alpha  \dot{\psi} \, \partial_t \left ( e^{-\alpha \Box} \psi
  \right ) \right.  - \nonumber \\
&~& \alpha \int_0^1 d\tau \, \left( e^{-\tau\alpha\Box}(1+4\xi^2 \alpha  \Box) \psi \right) \left(\Box e^{-(1-\tau)\alpha \Box} \psi \right) +  \nonumber\\
&~& \left. \alpha \int^1_0 d\tau \, 
\partial_t \left(e^{-\tau\alpha\Box}\left(1+4\xi^2 \alpha \Box\right) \psi \right )
\partial_t \left(e^{-(1-\tau)\alpha\Box}\psi \right ) \right ] \,.
 \nonumber \\
\end{eqnarray}

To make the connection with a diffusion-like pde, we 
introduce an auxiliary variable $r$, and define a new field as 
\be
\Psi(t,r)=e^{-r \alpha \Box}\psi(t)\,~,
\ee
which is a function of both time and the auxiliary variable $r$.  
By differentiating $\Psi(t,r)$ with respect to $r$, 
it is simple to see that $\Psi$ 
must satisfy the diffusion-like partial differential equation 
\be
\label{PDE}
\Box\Psi(t,r)=-\frac{1}{\alpha} \, \frac{\partial \Psi(t,r)}{\partial r} \,~.
\ee
Moreover, we can see that $\Psi(t,1) = e^{-\alpha \Box} \psi(t)$ and $\Psi(t,0) = \psi(t)$, 
hence, in terms of our new field $\Psi(t,r)$ the scalar field equation (\ref{ScalarField}) simply becomes 
\be
\label{Boundary}
\Psi(t,1) - 4\xi^2 \left[\frac{\partial \Psi(t,r)}{\partial r}\right]_{r=1} =
-\frac{\partial V(\Psi(t,0))}{\partial \Psi(t,0)} \,.
\ee
It is clear, therefore, that if $\Psi(r,t)$ is a solution to the pde (\ref{PDE}) 
which satisfies the boundary condition given by Eq.~(\ref{Boundary}) at all times, 
then $\psi(t)=\Psi(t,r=0)$, is a solution to the non-local field 
equations, and moreover, $H$ can also be determined through 
Eq.~(\ref{Fried}) (and in turn the behavior of the 
scale factor can be obtained). A 
complicating factor is that the Hubble rate $H$ appears 
in the d'Alembertian operator $\Box$, so $H$ and $\Psi(t,r)$ 
 must be solved for simultaneously.
This is however straightforward, since using the definition of $\Psi(r,t)$ and the 
diffusion equation (\ref{PDE}), the stress energy tensor 
(\ref{StressEnergy}) can be written purely in terms of 
$\Psi(t,r)$ for $r$ in the range
$0<r<1$, with no non-local operators appearing explicitly.
For the FRW case, the energy density, for example, is 
given purely in terms of $\Psi$ and its first time derivative as
\begin{eqnarray}
\rho &=&  \frac{\gamma^4 \lambda}{2} \left [ \Psi(t,0) \Psi(t,1) + 2V(\Psi(t,0)) \right.\nonumber \\
&-& 4\xi^2\alpha \dot{\Psi}(t,0) \dot{\Psi}(t,1) \nonumber \\
&-& \int^1_0 d\tau \, \left(\Psi(t, \tau)-4\xi^2 \frac{\partial \Psi(t,\tau)}{\partial \tau} \right)  \,
\frac{\partial \Psi(t,1-\tau)}{\partial \tau}  \nonumber \\
 &+& \left. \alpha \int^1_0  d\tau \, \left(\dot{\Psi}(t,\tau)-4\xi^2 \frac{\partial \dot{\Psi}(t,\tau)}{\partial \tau}\right) \dot{\Psi}(t,1-\tau) \right] \,. \nonumber \\
\end{eqnarray}

\subsection{Solving the diffusion-like pde as a boundary value problem}
\label{iterative}
One method of numerically solving the diffusion-like equation which gives rise to
a solution to the non-local field equations has been explored 
in previous studies \cite{ludaBoundary}.  In these works the pde 
(\ref{PDE})
is solved as a boundary value problem. This approach has the draw back
of being more restrictive in the solutions it can explore than an
initial value formulation, 
but nevertheless it has been highly successful and it is well
worth reviewing before we proceed to the initial value formulation.

\begin{figure}
\includegraphics[width=8.5cm]{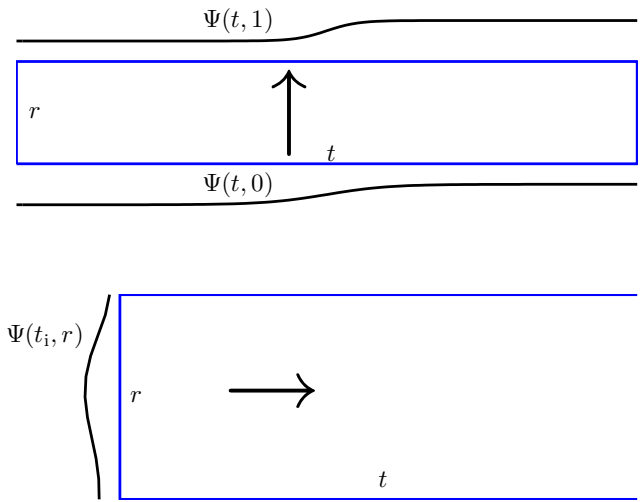}
\caption{\label{regiona} In the upper diagram we illustrate the method of solving the diffusion-like  equation as a boundary value problem. In the lower diagram, we show how the diffusion like equation is solved as an initial value problem.}
\end{figure}

The method attempts to find a solution to the pde in the
region  $0<r<1$ and $-\infty < t < \infty$, visualized in 
the upper panel of Fig.~\ref{regiona}. The method requires $\Psi(t=-\infty,r)$, and 
$\Psi(t=\infty,r)$ to be specified (i.e. the boundaries to be
fixed). The only obvious consistent data for these boundaries 
is for $\Psi$ to be constant for all values of $r$ and equal to field values  
corresponding to turning points of the full potential part of the action (\ref{Lag}), that is 
$V+\psi^2/2$.  At these points all the
infinite derivatives can be consistently set to zero, and the field is
stationary at the maximum or minimum of its potential. The other two
boundaries 
$\Psi(t,r=0)$ and $\Psi(t,r=1)$ are then assumed to be related 
through the condition given by Eq.~(\ref{Boundary}). 
Then a solution is found
using some form of relaxation/iterative procedure.  The simplest
example would be as follows:  initial trial functions for 
$\Psi(t,0)$ and $H(t)$ for $-\infty<t<\infty$ are set, making
sure that both asymptote to constants at $t=-\infty$ and $t=\infty$
which are consistent with the field being at an extremum of its
potential asymptotically in the past and future. The initial form 
of $\Psi(t,0)$ is then used as an
initial condition for numerical integration of the pde in the $r$ 
direction from $r=0$  to $r=1$. 
Once the initial trial function is 
integrated to $r=1$, a new initial function is determined by 
employing the boundary condition (\ref{Boundary}), the new 
trial function given by
\begin{eqnarray}
\Psi(t,0)_{\rm new} &=& \left(\frac{\partial V}{\partial \Psi(t,0)}\right)^{-1} (x) \,,
\nonumber \\
x &=& 4\xi^2 \left[\frac{\partial \Psi(t,r)_{\rm old}}{\partial r}\right]_{r=1} - \Psi(t,1)_{\rm old}  \,, 
\end{eqnarray}
where $(\partial V/\partial \Psi(t,0))^{-1}(x)$ denotes the inverse function of 
$\partial V/\partial \Psi(t,0)$.
A new trial function for $H(t)$ is also determined by 
employing $\Psi(t,r)_{\rm old}$ in Eq.~(\ref{Hdot}) or in Eq.~(\ref{Fried}). 
This procedure is then repeated until
convergence is achieved. Obviously variations on this procedure can
be used, and more sophisticated relaxation or other techniques
could be employed (also see \cite{volovich} and references therein for finding 
solutions iteratively not directly using the diffusion equation in the Minkowski space 
setting).

\subsection{Solving the diffusion equation as an initial value problem} 
\label{ivprob}
Despite its success, the boundary value formulation can only produce solutions 
with the particular 
initial and final conditions where all derivatives 
of the field tend to zero, and hence can only find solutions which evolve
from one extremum of the field's potential to another. 
Indeed, although interesting 
and physically relevant solutions can be obtained 
using this method, these requirements are rather restrictive.  
For example, we may wish to consider what happens to a non-local 
field which is 
initially at the maximum of its potential and then subjected to a small fluctuation and ask: what is the full range of possible evolutions that can result? Moreover, as we discuss in 
section \ref{Perts}, the boundary formulation when applied to perturbative 
equations is not able to produce physically interesting solutions.  
For a perturbation  
we would normally like to specify some initial conditions, for a local
field this is usually that the field initially be in the  
Bunch-Davis vacuum far inside the cosmological horizon, 
integrate the system forward in time and determine how 
the perturbation evolves. This is not possible within the framework of
the boundary value formulation.
  
An interesting question, therefore, is: can we solve a non-local cosmology 
as an initial value problem using the diffusion-like pde?  
Such a possibility would allow us to specify 
initial conditions beyond the restrictive ones discussed above, and help us 
to understand the freedom allowed in specifying initial conditions and 
hence the variety in the possible solutions to the field equations. 
This is the purpose of the present work.

The idea is illustrated in the lower panel of Fig.~\ref{regiona} and we 
now describe it in detail.  We 
specify some initial data $\{ \Psi(t_{\rm i},r), \dot{\Psi}(t_{\rm i},r) \}$, 
which in turn gives us the initial value $H(t_{\rm i})$ through 
Eq.~(\ref{Fried}).  We then use these initial functions 
together with the differential equation (\ref{PDE}) and 
the boundary condition (\ref{Boundary}) in the $r$ direction 
to evolve $\Psi(t_{\rm i},r)$ one time step forwards in time. 
The new $\Psi(t,r)$ is then used to calculate a 
new $H(t)$, and the procedure is repeated.  Continuing 
this procedure indefinitely ought to allow us to evolve the initial
data forwards in time. Naively everything seems as it should be, i.e., we
specify the two initial conditions we require at the initial time
boundary (the correct number as the equation is second order in time),
and have the required condition relating the $r=0$ and $r=1$ boundary.  
This boundary condition is unusual, but numerically its implementation
is not a problem. There are however two serious issues with this idea.
The first is that the system is ill-posed for generic initial data.  
This means that high frequency modes present in the initial function 
grow faster than lower frequency modes, and this leads to a numerical 
instability as any small numerical inaccuracies will show up as the 
introduction of high frequency modes which rapidly grow and swamp 
the solution we are looking for. This can be seen clearly 
for the case of 
constant $H$ where the diffusion equation can be solved exactly for linearised boundary conditions. 
The most general solution is given in Eq.~(\ref{DiffSolveGen}), and here we simply note that by studying that solution 
it can readily be seen that as the solution becomes more oscillatory in the $r$ direction, so its growth in time 
increases. It can also be seen that the Hubble rate $H$ has a damping effect decreasing the 
growth rate, but that sufficiently oscillatory solutions will always have a growing mode.
This seems a terminal problem, 
but we will see that in many circumstances of interest the instability 
can be suppressed by regularization methods and reliable solutions 
can still be found. 

The second issue is that the initial data  
$\Psi(t_{\rm i},r)$ and $\dot{\Psi}(t_{\rm i},r)$ cannot be arbitrary, but must 
also satisfy the boundary condition (\ref{Boundary}).  This is a more stringent condition than one might initially think 
as this condition implies an infinite number of other
conditions on the relationship between the function $\Psi(t_{\rm i},r)$ evaluated at 
$r=1$ and at $r=0$. These conditions result from successively differentiating (\ref{Boundary}). Let us then illustrate this for the case when $\dot{\Psi}(t_{\rm i},r)=0$ which is a consistent initial condition, and  
for convenience we take $\xi^2=0$ and $V(\psi) = -\psi^4/4$. 
Applying $\Box$ successively to the boundary condition (\ref{Boundary}) we find 
\begin{eqnarray}
\label{BoundaryMany1}
\Psi(t_{\rm i},1) &=& \Psi(t_{\rm i},0)^3  \,, \nonumber \\
\left[ \frac{\partial \Psi(t_{\rm i},r)}{\partial r}\right]_{r=1} &=& 
3 \,\Psi(t_{\rm i},0)^2  \left[\frac{ \partial \Psi(t_{\rm i},r)}{\partial r}\right]_{r=0}  \,,
\nonumber \\
\left[ \frac{\partial^2 \Psi(t_{\rm i},r)}{\partial r^2}\right]_{r=1} &=& 
3 \,\Psi(t_{\rm i},0)^2  \left[\frac{ \partial^2 \Psi(t_{\rm i},r)}{\partial r^2}\right]_{r=0} 
\nonumber \\
&~& + 18 \Psi(t_{\rm i},0) \, \left[\frac{ \partial \Psi(t_{\rm i},r)}{\partial r}\right]_{r=0}^2 \,, \nonumber \\
\end{eqnarray}
for the first three operations, where we have used the pde itself to replace $\Box$ with derivatives with respect to $r$.
We can of course continue this 
process indefinitely.  We now need to find a 
function of $r$ which satisfies an infinite set of constraints 
on its derivatives at its boundaries. In fact, this infinite set of
conditions is simply a reflection of the infinite number of 
constraints that would have to be imposed on the 
initial conditions for $\psi$, $\dot{\psi}$, etc, which we discussed in section 
\ref{cauchyproblem}.  
The task of finding this function $\Psi(t_{\rm i},r)$ is difficult, and we can 
only hope to
find a function that can approximately satisfy the conditions. We
propose to do this by computing the linear evolution of $\psi$ 
about a chosen value $A$ such that 
$\psi(t)= A+\phi(t)$ and hence $\Psi(t,r) = \exp(-r\alpha \Box)(A+\phi(t)) =
A + \Phi(t,r)$. We will see that this expansion will allow us to find
the profile in $r$ of $\Phi(t_{\rm i},r)$, the problem being much
simpler because the equation of motion is linearised.  This profile in turn 
gives us a profile for 
$\Psi(t_{\rm i},r)$ approximately. 
Let us then study the linear non-local equations first.
%

\section{Linear and linearised equations}
\label{lineareqns}
In this section we consider the situation when the equation of motion
(\ref{ScalarField}) is linear or it can be linearised by expanding
$\psi$ around a chosen point $A$ such that $\psi = A + \phi$. Our main
purpose is to aid our study of non-linear equations by linearising
them, but it should be noted that if the model was 
intrinsically linear this 
section furthers our understanding of these models in there own
right. 
Inserting $\psi=A+\phi$ 
into the action (\ref{Lag}) and expanding to quadratic  
order in $\phi$, it results that the equation of motion remains unchanged in terms of $\phi$,
\be
\label{eqm1}
(1+4\xi^2\alpha \Box) e^{-\alpha\Box} \phi = - \frac{dU(\phi)}{d\phi} \,,
\ee
but with a potential $U(\phi)$  given by
\be
\label{genpot}
U(\phi) = U_0 - c \, \phi - \frac{1}{2} m^2 \phi^2 \,,
\ee
where $U_0 = V(A)+A^2/2$, $c = -V'(A)-A$, $m^2 = -V''(A)$, 
a prime means differentiation with respect to $\psi$ and the quantities are evaluated at $\psi = A$. Let us now define the field $\Phi(t,r) = e^{-r \alpha \Box} \phi$ such that now we have the diffusion-like equation 
\be
\label{PDElinear}
\Box \Phi = -\frac{1}{\alpha} \frac{\partial \Phi}{\partial r} \,,
\ee
with the linear boundary condition
\be
\label{Boundarylinear}
\Phi(t,1) - 4 \xi^2 \left[\frac{\partial \Phi(t,r)}{\partial r}\right]_{r = 1} = m^2 \Phi(t,0) + c \,.
\ee
The diffusion-like equation can be solved by separation of variables using
\be
\label{SepSol}
\Phi(t,r) = \phi(t) g(r) + h(r) \,.
\ee
Substituting into Eq.~(\ref{PDElinear}) we obtain that the functions $g$ and $h$ must be of the form $g = e^{\alpha\omega^2 r}$, $h = b\,e^{\alpha\omega^2 r}/\omega^2 + k$ while $\phi$ satisfies the local equation
\be
\label{LocalField}
\Box \phi = -\omega^2 \phi -b \,,
\ee
where $k$, $\omega^2$ and $b$ are constants. By imposing $\Phi(t,0) = \phi(t)$ we obtain $k = -b/\omega^2$
and by making use of the boundary condition (\ref{Boundarylinear}) we obtain $b = c \,\omega^2/ (m^2-1)$ and the characteristic equation
\be
\label{characteristic1}
e^{\alpha\omega^2} (1-4\xi^2\alpha\omega^2) = m^2 \,.
\ee
We note at this point that to solve for $\phi$ it is only required to
solve a local equation with parameters $\omega^2$ and $b$. In fact, as it
has been done in previous literature, we could have postulated
(\ref{LocalField}), substituted it into the equation of motion
(\ref{eqm1}) and we would have obtained the same expression for $b$
and the characteristic equation (\ref{characteristic1}) 
without any mention of the diffusion-like equation. 
Indeed, the diffusion-like equation is only required when we move to
the 
non-linear boundary conditions as in that case we cannot localize the
field. 
This section, however, serves to illustrate that the two methods are 
equivalent in the linear case.

Let us consider in detail the roots of the characteristic equation (\ref{characteristic1}). In the case $\xi^2=0$, we have for $m^2 \neq 0$:
\be 
\label{aw2a}
\alpha \omega^2 = \ln m^2  \pm 2n \pi \, i \,,
\ee
and for $m^2 = 0$:
\be
\label{aw2b}
\alpha \omega^2 = -\infty \,,
\ee
where $n$ is an integer. As we will see below, this means that in this case the dynamics of a single,
massive, non-local field can be rewritten in terms of local field
equations for at most one real field together with an infinite
set of complex fields. We will see 
later that the complex fields can be coupled with their respective
complex 
conjugates to form two physical real fields which make up a
``quintom". 
For a free field,  $m^2 = 0$, the dynamics does not admit a local description.

For the case $\xi^2 \neq 0$, one obtains
for $m^2 \neq 0$: 
\be 
\label{aw2c}
\alpha \omega^2 = \frac{1}{4\xi^2} + {\rm LambertW}\left(n,x\right) \,,
\ee
where $x = -m^2 \exp(-1/4\xi^2)/4\xi^2$, 
and for $m^2 = 0$:
\be
\label{aw2d}
\alpha \omega^2 = \frac{1}{4\xi^2} \,,
\ee
and $n$ is again an integer. 
Hence, provided $m^2 \neq 0$, a real root exists for 
$x  \gtrsim -0.37$ when $n = 0$ and for $-0.37<x<0$ when $n=-1$, otherwise, the roots are complex. As we will see, it follows 
that the dynamics allows for a description in terms of at most two real fields and an infinite 
number of complex fields. 
For linearised models, the approximate  
local description resulting from an expansion at a given point $A$ in the potential 
is very useful as it gives us an understanding about the possible evolution of the field, 
for example, near an extremum of the potential. However, to understand
how the solutions evolve beyond any linearised regime, or how
they interpolate between say a maximum and a minimum one is required to go beyond the linear 
expansion into the non-linear regime of the evolution. To do this is one of the main aims 
of our work and we show numerical results for two string models in Sec.~\ref{applications}.

\subsection{The single real field solution}
\label{singlefield}
Before moving to the non-linear case, it is instructive 
to see how in the linear
case a correspondence can be made between the non-local dynamics and local dynamics in terms of canonically normalized fields, phantom fields and 
quintoms, which we defined in the introduction. First we will discuss 
the situation when only a single real root to the characteristic equation Eq.~(\ref{characteristic1}) is considered. 
We are interested in real solutions to the non-local equations, and so we will only be 
interested in situations in which $\phi$ is real. Inserting $\psi = A + \phi$ into 
the expression for the stress energy tensor (\ref{StressEnergy}) and employing Eq.~(\ref{LocalField}), 
the integrals can now 
be evaluated and we find that eq.~(\ref{StressEnergy}) can be written in terms of a real rescaled field
\be
\varphi = |B| \, \phi \,,
\ee
with $B^2 = \gamma^4 \lambda \alpha \left[1-4\xi^2(1+4\alpha\omega^2)\right] e^{\alpha\omega^2}$
such that
\be
T_{\mu\nu}^{\rm local} = s g_{\mu\nu} \left(\frac{1}{2} \partial_{\sigma}\varphi
\partial^{\sigma}\varphi + W(\varphi) \right) - s \partial_\mu \varphi \partial_\nu \varphi \,,
\ee
where $s$ is the sign of $B^2$ and the potential for the canonically normalized field is
\be
W(\varphi) = W_0 - s \tilde{b} \varphi - s \frac{1}{2} \omega^2 \varphi^2 \,,
\ee
where $\tilde{b} = |B| \,b$,  and 
\be
W_0 = \gamma^4 \lambda \left[ U_0 -\frac{1}{2} \left(\frac{b}{\omega^2}\right)^2 \left(1+ \alpha\omega^2 \, m^2  - e^{\alpha\omega^2} \right) \right] \,.
\ee
Comparing these equations to Eq.~(\ref{RealAct}), we see that when $B^2$ is positive 
$\varphi$ represents a canonically normalized field, while
when it is negative it represents a phantom field. 
It is clear, therefore, that $\phi$ is a solution to the
non-local field equation (\ref{eqm1})
if $\varphi$ is a solution to the local Einstein and scalar field equations 
\begin{eqnarray}
\Box\varphi &=&  -\omega^2 \varphi - \tilde{b} \,,  \nonumber\\
G_{\mu \nu} &=&\frac{8\pi}{\mpl^2} \, T^{\rm local}_{\mu \nu} \,~.
\end{eqnarray}
This remarkable simplification 
has been explored by a number of authors (for example Ref. \cite{lidsey,ludaLinear, koshelev,calcagni1}, the ansatz Eq. (\ref{LocalField}) was first proposed in Ref. \cite{biswas}).
In particular,  in Ref.~\cite{lidsey} a general non-local system with a
hilltop potential of the form (\ref{genpot}), but with $c=0$, 
was rewritten in the manner presented 
here in terms of a local field and it was found that in order to act as an inflationary potential satisfying 
the current observational constraints, the parameters must be $W_0 \approx 3.7 \times 10^{-10}$ and 
$\omega^2 = 3 \times 10^{-6}$, in Planck units. We will use these values below when we perform 
numerical integrations.

\subsection{Multiple fields}
\label{multiplefields}
In general the 
characteristic equation (\ref{characteristic1}) does not have a single root, 
but as we have seen, 
a set of roots $\omega^2_i$.
Moreover, since the non-local field
equation (\ref{LocalField}) is linear, 
the most general solution to this equation is given by
\be
\phi=\sum_i \phi_i\,~,
\ee 
where $\phi_i$ are fields satisfying individually 
\be 
\label{gendalembert}
\Box \phi_i = -\omega^2_i \phi_i - b_i \,,
\ee
where we now require the constants $b_i$ to satisfy the expression
\be
\label{bcondition}
\sum_i \frac{b_i}{\omega_i^2}  = \frac{c}{m^2-1}\,.
\ee
We see that for this general solution we have  
\be
\label{GenSol}
\Box\phi= -\sum_i \left( \omega^2_i \phi_i + b_i\right)   \,~,
\ee 
and inserting this into the stress-energy 
tensor (\ref{StressEnergy}) and performing the integrations, 
we arrive at: 
\begin{eqnarray}
\label{StressGen}
T^{\rm local}_{\mu \nu} &=&  g_{\mu \nu}  \left( \frac{1}{2} \sum_i
\partial_{\sigma}\varphi_i \partial^{\sigma}\varphi_i + W(\varphi_i) \right)
\nonumber \\
&~& - \sum_i  \partial_\mu \varphi_i
\partial_\nu \varphi_i \,~.
\end{eqnarray}
where now the scalar potential involving all the fields is 
\be
W(\varphi_i) = W_0 - \sum_i \tilde{b}_i \varphi_i - \frac{1}{2} \sum_i \omega_i^2 \varphi_i^2 \,,
\ee
with $W_0$ given by
\begin{eqnarray}
W_0 &=& \gamma^4 \lambda U_0 -
2 \gamma^4 \lambda \sum_i \sum_{j>i} \frac{b_i b_j}{\omega_i^2 \omega_j^2} (1-m^2) - \nonumber \\
&~& \gamma^4 \lambda \sum_i \left(\frac{b_i}{\omega_i^2}\right) \left(1+ \alpha\omega_i^2 \, m^2
- e^{\alpha\omega^2} \right) \,.
\end{eqnarray}
The remarkable fact is that all cross terms amongst the fields  
vanish, and Eq.~(\ref{StressGen}) 
is simply the stress energy tensor for a collection of local 
(but in general complex) fields $\varphi_i = B_i \phi_i$, with
$B_i^2 = \gamma^4 \lambda \alpha \left[1-4\xi^2(1+4\alpha\omega_i^2)\right] e^{\alpha\omega_i^2}$. 
This has been discussed in \cite{ludaLinear}, and a very similar analysis, 
starting with the Lagrangian rather than the field equations, 
was presented in \cite{koshelev}. 
We note that previous works which considered multiple roots of the characteristic equation took $c = 0$ in 
Eq.~(\ref{genpot}), here we have also included the linear term in the potential.

\subsection{Quintoms from pairs of complex fields}
\label{quintoms}
For real roots to the characteristic equations we have seen that we can 
rewrite the resulting local fields as   
canonically normalized or phantom fields. However, complex roots, 
when viewed in isolation, would lead to complex local fields 
and an unphysical evolution (i.e., $H$ and consequently  
the scale factor $a$ would be complex).    
It is important to understand, however, that 
they can be combined in such a way as to produce real fields \cite{koshelev}.  
This follows because the
complex roots of Eq.~(\ref{characteristic1}) always come in pairs which are 
the complex conjugates of one another. For every root $\omega^2_i$, it can readily be
verified that
 its complex conjugate  $(\omega^2_i)^*$ is also a root, 
moreover, the associated 
 $b_i$ and $b_i^*$ of these roots can also be taken to be the  
complex conjugates of one another. 
For convenience we relabel pairs of complex roots such that 
$\omega^2_m$ is the complex conjugate of $\omega^2_{-m}$ where $m$ is a 
positive integer, and if real roots exists we label them
$\omega^2_r$ (note that for certain parameter choices $m = n$, but this is 
not generically true, for example, in the case $\xi^2\neq 0$ there can
exist a complex root even for $n=0$ or $n= -1$).

Considering the pair of complex conjugate 
roots, Eq.~(\ref{LocalField}) implies in turn that 
$\varphi_m$ and $\varphi_{-m}$ are also the complex conjugates of 
one another, under the assumption that the initial conditions from 
which the fields 
$\varphi_{m}$ follow are the complex conjugates of the initial 
conditions for the fields $\varphi_{-m}$. The combination 
$\varphi_m+\varphi_{-m}$ is 
therefore real and satisfies local field equations. 
To understand the dynamics further, let us split the complex fields 
into their real and imaginary parts thus:  
\begin{eqnarray}
\varphi_m &=& \frac{1}{\sqrt{2}} (\chi_m + i \sigma_m) \,, \\
\varphi_{-m} &=& \frac{1}{\sqrt{2}} (\chi_m - i \sigma_m) \,,
\end{eqnarray}
where both $\chi_m$ and $\sigma_m$ are real valued fields.
The real field which follows from adding the conjugate fields 
then simply becomes $\varphi_m + \varphi_{-m} = \sqrt{2} \chi_m$. 
We also split the roots such that, $\omega_m^2=\alpha_m^2 + 
i \beta_m^2$. 
Rewriting the local stress energy tensor
(\ref{StressGen}) using this splitting, we find that it becomes 

\begin{eqnarray}
\label{StressLocalQuintom}
T^{\rm local}_{\mu \nu} &=&
 g_{\mu \nu} \left[ W(\varphi_r,\chi_m,\sigma_m) +  \frac{1}{2} \sum_r s_r\partial_\mu \varphi_r \partial_\nu \varphi_r + \right. \nonumber \\
 &~& \left.
\frac{1}{2} \sum_m \left( \partial_{\sigma}\chi_m  \partial^{ \sigma} \chi_m - 
\partial_{\sigma}\sigma_m  \partial^{ \sigma} \sigma_m \right) \right] - \nonumber \\
&~& \sum_r s_r \partial_\mu \varphi_r \partial_\nu \varphi_r  - 
\sum_m \left( \partial_\mu \chi_m \partial_\nu \chi_m - 
\partial_\mu \sigma_m \partial_\nu \sigma_m \right) \,.  \nonumber \\
\end{eqnarray}
where the general scalar potential is
\begin{eqnarray}
W(\varphi_0,\chi_m,\sigma_m) &=& W_0 -  \sum_r s_r \left( \tilde{b}_r \varphi_r - \frac{1}{2} \omega_r^2 \varphi_r^2\right) - \nonumber \\
&~&
\sum_m \left( \tilde{p}_m \chi_m - \tilde{q}_m \sigma_m \right) - \nonumber \\
&~& 
 \frac{1}{2} \sum_m \left[ \alpha_m^2 (\chi_m^2-\sigma_m^2) 
- \beta_m^2 \chi_m \sigma_m
\right] \,, \nonumber \\
\end{eqnarray} 
where we have defined $\tilde{b}_m = (\tilde{p}+i\tilde{q})/\sqrt{2}$.
We can also write 
\begin{eqnarray}
\label{ScalarQuintoms}
\Box \varphi_r &=& -\omega_r^2 \varphi_r - \tilde{b}_r \,, \nonumber  \\
 \Box\chi_m &=&  
-\alpha_m^2 \chi_m^2 + \beta_m^2 \sigma_m -\tilde{p}_m \,, \nonumber \\
\Box \sigma_m &=& -\beta_m^2 \chi_m -\alpha_m^2\sigma_m - \tilde{q}_m \,.
\end{eqnarray}
Considering $G_{\mu \nu} = 8 \pi T^{\rm local}_{\mu \nu}/\mpl^2$ together with 
Eqs.~(\ref{ScalarQuintoms})
we see that we have succeeded in writing the most general local field
equations in terms of purely real fields.  
The field $\chi$ is a regular canonically normalized field whereas $\sigma$ is a phantom as its kinetic energy comes with the wrong sign. 
The most general set of local fields which solve the non-local 
equations are therefore 
given by at most two real canonically normalized fields or phantom fields, and an infinite 
set of quintoms.

So far our work on the correspondence construction has been completely 
general, 
but as our primary interest is in 
cosmological dynamics, we once again specialize to the FRW metric. 
The field equations become the Friedmann equation (\ref{Fried}) and the 
Raychaudhuri equation (\ref{Hdot}) where 
\be
\rho = \frac{1}{2} \sum_r s_r \dot{\varphi}_r^2 + \frac{1}{2} \sum_m \left( \dot{\chi}_m^2 - \dot{\sigma}_m^2 \right) 
+ W(\varphi_r,\chi_m,\sigma_m) \,,
\ee
and
\be
P = \frac{1}{2} \sum_r s_r \dot{\varphi}_r^2 + \frac{1}{2} \sum_m \left( \dot{\chi}_m^2 - \dot{\sigma}_m^2 \right) 
- W(\varphi_r,\chi_m,\sigma_m) \,,
\ee
are the total energy density and pressure, respectively.

These field equations can easily be solved for numerically, so 
long as we include only a finite number of fields in our sum. 
This is consistent since setting $\chi_m=\sigma_m=\dot{\chi}_m=\dot{\sigma}_m=0$ 
initially, and implies that this state will be maintained indefinitely so
long as $b_m=0$, hence, we are allowed to make 
only a finite number of quintoms dynamically active. 
Specifying initial data to 
solve the linear non-local cosmology as an initial value problem, 
therefore, reduces to 
specifying the initial magnitude and velocity of each field in 
our infinite set.

At this point it is worth recalling a recent analysis
in Ref.~\cite{ivp}, which sought to understand the initial conditions required to
solve linear non-local theories as an initial value problem.  Though
the analysis presented there is more general than the one given here (but less complete in cosmological settings), it is clear that the two are equivalent as it was found there that each pole 
of a suitably defined propagator (equivalent to the zeros of our characteristic equation), requires two 
initial conditions (equivalent to our initial field position and velocity).

\section{Diffusing linear non-local cosmology}
\label{difflinear}

For our linear theory the diffusion-like equation can be solved in time 
once we specify initial 
functions $\Phi(t_{\rm i},r)$ and $\dot{\Phi}(t_{\rm i},r)$
which satisfy the boundary condition (\ref{Boundarylinear}).  
But which class of functions are acceptable for the initial data? 
We found in the previous section that when only one field is present, 
the general form of $\Phi(t,r)$, Eq. (\ref{SepSol}), is
\be
\label{Phi1}
\Phi(t,r) = \phi(t) e^{\alpha\omega^2 \, r} + \frac{b}{\omega^2}\left(e^{\alpha\omega^2 \, r} -1\right) \,.
\ee
Hence, a suitable initial profile can simply be taken to be
\be
\Phi(t_{\rm i},r) =  k - C \, e^{\alpha\omega^2 r}  \,,
\ee
where $k = -b/\omega^2$, $C$ is an arbitrary constant and $\alpha\omega^2$ is a solution of the characteristic equation (\ref{characteristic1}).
We have already studied the possible values for $\omega_i^2$, 
and we found that they can have at most two real values and an 
infinite number of complex values. We have also learnt that for
each complex root, the complex conjugate value is also an acceptable 
root,
hence, we can add or subtract $C \, e^{\alpha\omega^2r}$ and 
$(C e^{\alpha\omega^2 r})^*$ in order that $\Phi(t_{\rm i},r)$ and $\dot{\Phi}(t_{\rm i},r)$ are real. 
Therefore, the initial data can be a linear combination of the following:
\begin{eqnarray}
\label{ics}
\Phi_1(t_{\rm i},r) &=& 2k_R - 2|C| \, e^{\alpha  \omega^2_R r} 
\cos\left(\theta +\alpha \omega^2_I \, r \right)\,,   \nonumber\\
\Phi_2(t_{\rm i},r) &=& 2k_I - 2|C| \, e^{\alpha\omega^2_R r} 
\sin \left(\theta + \alpha\omega^2_I \, r \right)  \,.  
\end{eqnarray}
where the indices $R$ and $I$ represent the real and imaginary parts of $\omega^2$ and $k$.
To confirm that these functions do indeed represent suitable initial data, we must calculate all the infinite constraints implied by the boundary condition (\ref{Boundarylinear}). Let us do this for the linearised model defined by 
$\xi^2=0$, 
$V(\psi) = -\psi^4/4$ and $\psi = 1+ \phi$ (c.f. Eq.(\ref{BoundaryMany1})).  Because of linearity this is simple and we find
\begin{eqnarray}
\label{BoundaryMany}
\Phi(t_{\rm i},1) &=& 3 \Phi(t_{\rm i},0)  \,, \nonumber \\
\left[\frac{\partial \Phi(t_{\rm i},r)}{\partial r}\right]_{r=1} &=& 3 \left[\frac{ \partial
      \Phi(t_{\rm i},r)}{\partial r}\right]_{r=0}  \,, \nonumber \\
\left[\frac{\partial^2 \Phi(t_{\rm i},r)}{\partial r^2}\right]_{r=1} &=& 3 \left[\frac{ \partial^2
      \Phi(t_{\rm i},r)}{\partial r^2}\right]_{r=0}  \,,
\end{eqnarray}
and similar for higher order derivatives. It is straightforward to 
verify that the trial functions (\ref{ics}) satisfy these boundary conditions.
One can easily verify that even for the more complex cases when $\xi^2 \neq
0$, the corresponding boundary conditions are still satisfied to all
orders in the linear approximation. 

Before we move to the non-linear non-local case it is
worth pointing out that for constant $H$ the diffusion equation with
linear boundary conditions can be solved exactly. Here we give the most general solution 
that satisfies the boundary condition (\ref{Boundarylinear}) for the case of $b=0$,  
\begin{eqnarray}
\label{DiffSolveGen}
\Phi(t,r) = 
\sum_{\omega^2} &~& \left\{ A_{\omega^2} e^{R_+ t + \alpha\omega^2_R r}\left [\cos(\alpha\omega^2_I r)\cos(I_1 t)  \right. \right. \nonumber \\ 
&~& \left. \left. - \sin(\alpha \omega^2_I r)\sin(I_1 t)\right] \right. \nonumber \\
&~& + \left. B_{\omega^2} e^{R_- t + \alpha \omega^2_R r} \left [\cos(\alpha \omega^2_I r)\cos(I_2 t) \right. \right. \nonumber \\ 
&~& \left. \left. - \sin( \alpha \omega^2_I r)\sin(I_2 t)\right ]\right\}\,,
\end{eqnarray}
where $A_{\omega^2}$ and $B_{\omega^2}$ are arbitrary constants and $R_{\pm}= (1/2){\rm Re}\left(-3H \pm \sqrt{9 H^2 + \alpha\omega^2}\right)$ and $I_{\pm}= (1/2){\rm Im}\left(-3H \pm \sqrt{9 H^2 + \alpha\omega^2}\right)$.
This highlights the ill-posed nature of our problem, since
it can be verified that as $\omega^2_I$ increases so that the solution becomes more oscillatory 
in $r$ (which generically occurs for roots to the characteristic equation with larger and larger $n$ values) 
the growth rate of the growing mode also increases.

\section{Diffusing non-linear non-local cosmology}
\label{diffnonlinear}
Like in the linear case, if we are to solve the full non-linear equations as an initial value problem using the diffusion-like equation, then we must specify an initial 
function which satisfies all the infinite conditions that we started to write down for the 
$\xi^2=0$, $V(\psi) = -\psi^4/4$ case in Eq.~(\ref{BoundaryMany1}).  In general this is very difficult,  however, we know that when close to the point about which we made our linearised construction, an approximate initial function takes the form 
\be
\Psi(t_{\rm i},r) = A+\Phi(t_{\rm i},r) \,,
\ee
where $\Phi(t_{\rm i},r)$ is one of the functions given in Eq.~(\ref{ics}).  One can 
verify that this trial function approximately satisfies the boundary conditions (\ref{BoundaryMany1}) and 
that the closer to the expansion point $A$, the better the approximation, and 
therefore it will suit as an initial condition for a numerical evolution.

\section{Applications and numerical integration}
\label{applications}
We have implemented the procedure described in the previous sections to compute the time evolution of a field $\psi$ for two string models. They have in common that the general form of the potential  in action (\ref{Lag}) can be written as 
\be
V(\psi) = \Lambda -\frac{1}{p+1}\, \psi^{p+1} \,.
\ee
Hence, the parameters $c$ and $m^2$ in the linearised potential (\ref{genpot}) read
\begin{eqnarray}
c = A^p -A \,, \hspace{1cm} m^2 = pA^{p-1} \,.
\end{eqnarray}

\subsection{$p$-adic string}
\begin{figure}
\includegraphics[width=8.5cm]{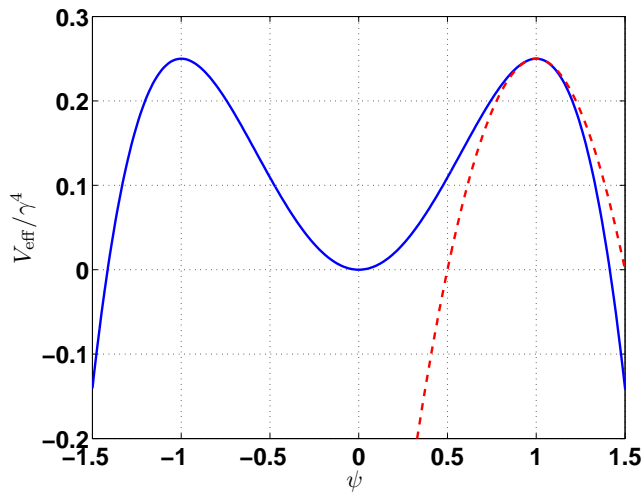}
\caption{\label{veff1} The form of the effective potential for $\psi$ in the $p$-adic string with $p = 3$. The solid line and dashed lines represent the full potential and the quadratically expanded potential around the hilltop, respectively.}
\end{figure}

In the $p$-adic string $\Lambda = 0$ and the parameters in action (\ref{Lag}) correspond to 
$\lambda = 1$, $\xi^2 = 0$, and 
\begin{eqnarray}
\alpha = \frac{\ln p}{2 m_s^2} \,, \hspace{1cm} 
\gamma^4 = \frac{m_s^4}{g_s^2} \, \frac{p^2}{p-1} \,,
\end{eqnarray}
where $g_s$ and $m_s$ are the string coupling and string scale, respectively.
Expanding the action (\ref{Lag}) to first order in $\Box$ we can see that it is
equivalent to the action of a field with a non-canonical kinetic term
\be
\label{cutaction1}
S_\psi = \int d^4 x \sqrt{-g} 
\left( \frac{1}{2}\gamma^4 \alpha \psi \Box \psi - V_{\rm eff}(\psi)   \right) \,,
\ee
where the effective potential is 
$V_{\rm eff}(\psi) = \gamma^4(V(\psi) + \psi^2/2)$ which we illustrate in Fig.~\ref{veff1} for $p$ = 3.
Including the higher order terms in the derivatives does not change
the  potential part of the action, but affects how the field evolves on this
potential via the additional kinetic terms. 

In our numerical integration of the equations of motion, we have set the initial data by expanding the solution near the 
hilltop, $A=1-\epsilon$, and have therefore fixed 
\begin{eqnarray}
\label{profile1}
\Psi(t_{\rm i},r) &=& 1 - \epsilon \, e^{\alpha\omega^2_R \, r} 
\cos(\alpha\omega^2_I r) \,, \\
\label{profile2}
\dot{\Psi}(t_{\rm i},r) &=& 0 \,,
\end{eqnarray}
with $\epsilon = 10^{-3}$, $\omega^2$ is a root given by Eq.~(\ref{aw2a}),
$\alpha = \ln(p (1-\epsilon)^{p-1})/(9\times10^{-12})$ with $p=3$, 
and $\gamma^4 = 1.48\times10^{-9}$ in Planck units.
We are now ready to evolve the diffusion
equation (\ref{PDE}). We have implemented a number of finite difference
schemes and employed the method of lines, in which the $r$ direction
is discretised and ordinary differential equation routines were used to
integrate the resulting equations in $t$. The results are robust to the
methods chosen.

As we mentioned before in Sec.~\ref{ivprob}, the numerical integration
of the diffusion equation is plagued with instabilities. This is
particularly true when $H$ is small or negative, since we saw that
large $H$ decreases the rate at which high frequency modes grow. 
We therefore
require to smooth the profile $\Psi(t,r)$ in $r$ at each time 
step as it evolves, in order to remove the spurious oscillations resulting from
numerical errors. These usually enter the evolution on the 
scale of the finite grid space in the $r$ direction as this
represents the highest frequencies permitted. We perform this
smoothing in a number of ways in order to ensure the results are not
overly dependent upon it. Since oscillations appear at high frequencies, one
possibility is to take an average over a number of points around each
point in $r$, in order to smooth the function on small scales but leave
it unchanged on large scales. This can prove difficult
to do fairly at points close to the boundaries so a procedure with similar
results is to take a subset of points $r_*$, evaluate $\Psi(t,r_*)$ at
those points and perform a spline fitting to reconstruct the data
in-between.  This procedure proves to work very well until $H$ becomes negligible. Another possibility, given what we have learnt about the form of the profile and the properties of the roots to the characteristic equation, is to fit the profile to a combination of Eqs.~(\ref{ics}).
Moreover, it can be verified that the real parts of the roots vary substantially between $\psi = 1$ and $\psi=0$, however, the imaginary parts are a constant in the case of $\xi^2 = 0$.
Considering this, we find that the profile at a given time step can be well fitted by
\be
\label{psifit}
\Psi_{\rm fit}(t,r) = c_1 + c_2 \, c_3^r 
\cos(c_4 + \alpha\omega^2_I r) \,,
\ee
where $c_1$, $c_2$, $c_3$ and $c_4$ are constants to be determined at
each time step. Again this theoretically motivated 
method proves highly successful. We have
also tried fitting at each time step to other functions with 
smooth properties such as polynomials. Though with less success, the results 
show significant improvement over no smoothing at all. All these method are
forms of regularization, which generically one must do when dealing
with ill-posed problems.  In fact, our methods are the most naive and  
simplest possibilities, though they appear to work well in the setting presented in this article. In future work we hope to return to the issue of regularization methods and utilize more sophisticated techniques.

We use the two methods which work best (the spline and the
theoretically motivated fitting) to cross check the results 
and we indeed find that they provide the same solutions in their 
respective region of
validity. Moreover we keep track of
the accumulated difference $\delta = \int dt \int dr |\Psi_{\rm
  numerical}-\Psi_{\rm fit}|$ and we choose the size of the grid in
$r$ and the time step such that this quantity is small. As a final
check on the validity of our solutions, we substitute the solution we
find back into a finite difference scheme (which can differ from the
one we use to generate results), and check the finite difference
equation is satisfied for all $r$ and $t$.

\begin{figure}
\includegraphics[width=8.5cm]{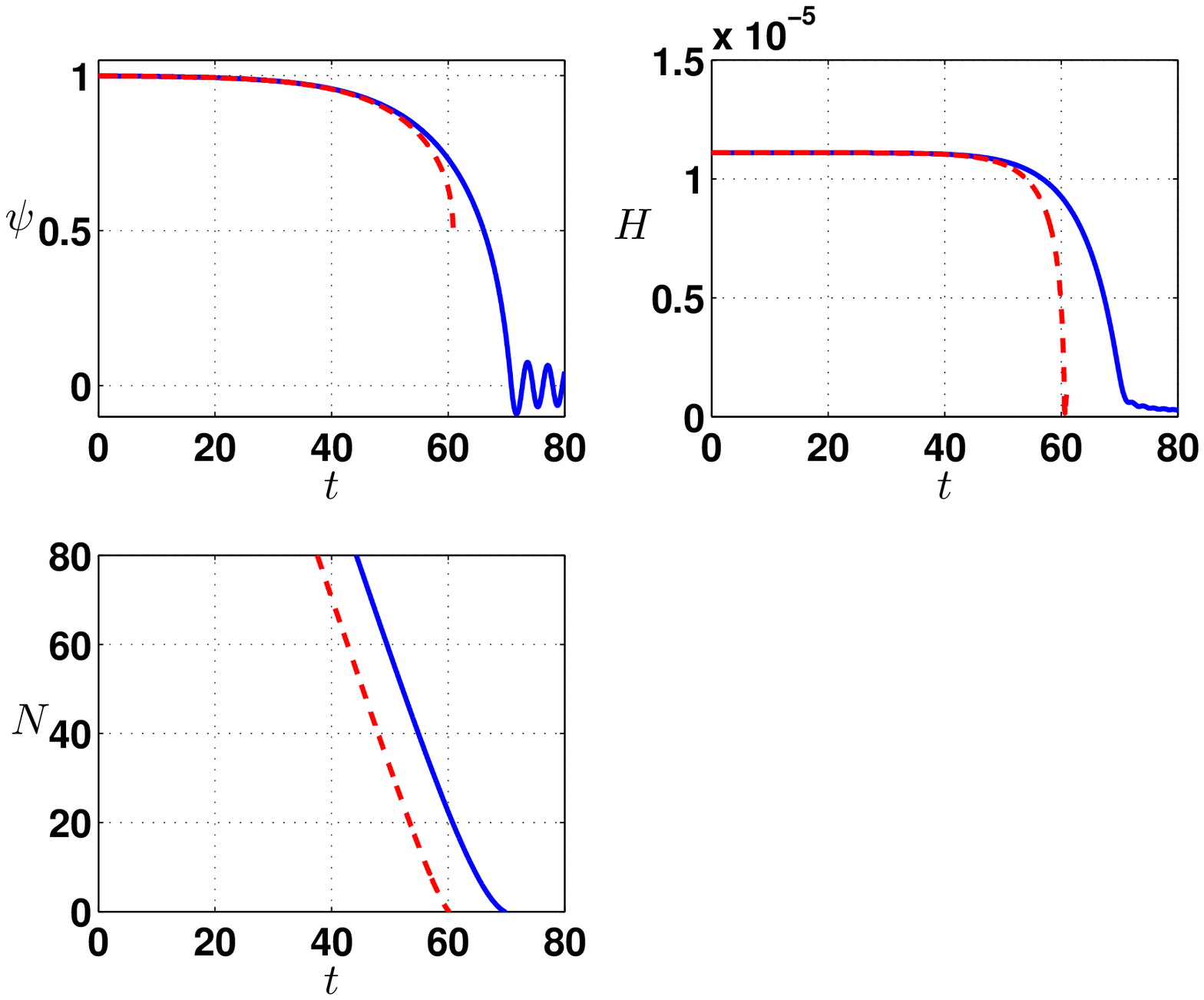}
\caption{\label{padicres1} Upper left panel: Evolution of the field $\psi$ in the $p$-adic string, $\xi^2=0$, $p=3$, where $\psi$ is a single real field, $n=0$, and initially $A = 1-\epsilon$. The solid and dashed lines represent the non-linear and linear evolution, respectively. Upper right panel: The corresponding evolution of the Hubble rate. Lower left panel: The number of $e$-folds before the end of inflation.}
\end{figure}

The results of the numerical integrations are shown in Fig.~\ref{padicres1} where the time variable was rescaled with respect to the true cosmic time by $t \rightarrow t/\sqrt{\alpha}$ for ease of computation.
Setting the initial profile using the $n=0$ root to the characteristic equation (i.e. $\omega_I^2=0$),  
we see that the evolution resulting from the 
linear expansion of the potential around the hilltop leads the Hubble 
rate to vanish when, in the language of the canonically normalized field, 
the potential $W(\varphi)$ 
cancels out the contribution from the kinetic term $\dot{\varphi}^2/2$.
Once $H$ becomes zero, a recollapse occurs and we do not follow the dynamics any further.
Evolving the field in the full non-linear potential shows that the
field settles at the minimum at $\psi =  0$ performing damped
oscillations and the Hubble rate, $H$, slowly decreases with time. 

This behavior is to be expected because if we make a local approximation 
to the dynamics by expanding about
the hilltop $\psi=1$ the result for the 
$n=0$ root to the characteristic equation 
is that the dynamics are equivalent to those
of a local canonically normalized field evolving on a hill-top 
potential (this is the result of \cite{lidsey}). On the other hand, 
if we make an expansion at a small value of $\psi$ (note
that though no expansion is possible at $\psi=0$, we can make the
expansion arbitrarily close to $\psi=0$), the dynamics for the $n=0$ root 
are equivalent to
a canonically normalized field evolving in a potential minimum made up
of a quadratic term together with a small linear piece. The expected
dynamics is for the field to oscillate about the minimum of this potential 
and to decay towards it.  The non-linear numerical integration appears to
interpolate between these two behaviors. 
In Fig.~\ref{padicres1}, we also show the number of $e$-folds before the end of inflation, $N$, and we can clearly see that when $N \gtrsim 60$, the linear and non-linear evolutions are practically indistinguishable. This observation suggests that the computation of the scalar power spectrum using only the linear approximation in 
Ref.~\cite{lidsey} is sufficiently accurate.
It is also interesting to compare our solution with the solution recently
found for the light-like $p$-adic tachyon in a linear dilaton background which
also evolves to the minimum of its potential \cite{Hellerman:2008wp}.

\begin{figure}
\includegraphics[width=8.5cm]{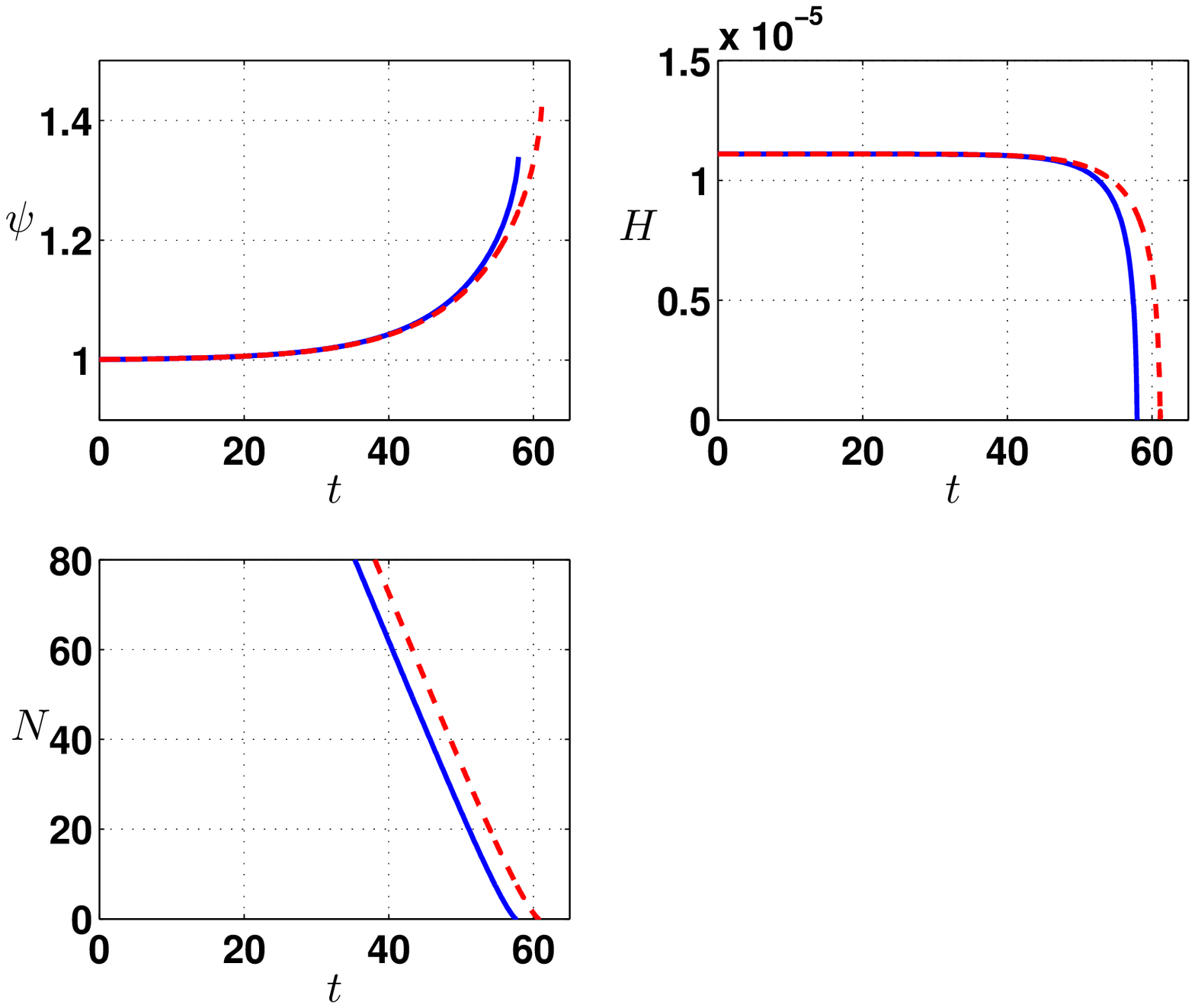}
\caption{\label{padicres1a} Upper left panel: Evolution of the field $\psi$ in the $p$-adic string, $\xi^2=0$, $p=3$, where $\psi$ is a single real field, $n=0$, and initially $A = 1+\epsilon$. The solid and dashed lines represent the non-linear and linear evolution, respectively. Upper right panel: The corresponding evolution of the Hubble rate. Lower left panel: The number of $e$-folds before the end of inflation.}
\end{figure}

In Fig.~\ref{padicres1a} we show the evolution of the same quantities as in Fig.~\ref{padicres1} but with the field starting its evolution at 
$A = 1+\epsilon$. Clearly the linear evolution is identical to the previous situation but the non-linear dynamics is different as there is no minimum to stabilize the field on the right hand side of the hilltop. Though the solution is not of particular interest, it highlights the utility of our method as opposed to a boundary value formulation as in that approach the field is only allowed to evolve between two extrema.

\begin{figure}
\includegraphics[width=8.5cm]{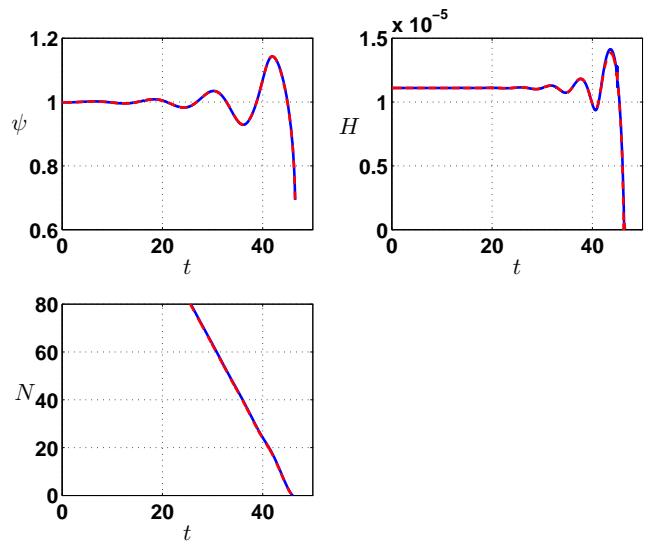}
\caption{\label{padicres2} Upper left panel: Evolution of the field $\psi$ in the $p$-adic string, $\xi^2=0$, $p=3$, where $\psi$ is the first quintom, $n=1$. 
The solid and dashed lines represent the non-linear and linear evolution, respectively. Upper right panel: The corresponding evolution of the Hubble rate. Lower left panel: The number of $e$-folds before the end of inflation.}
\end{figure}

With $n=1$, the first quintom, the field performs growing oscillations
around the hilltop as can be seen in Fig.~\ref{padicres2}. These
oscillations are mimicked by the Hubble rate and eventually they are
so large that cause this quantity to vanish. This occurs very near the
hill top, hence, the linear and non-linear evolutions are practically
indistinguishable. Once $H$ goes to zero we do not follow the
evolution any further as the universe recollapses at this
point. Although a re-expansion is a possibility, such cyclic
dynamics is more difficult to deal with as negative values of $H$
lead to even greater numerically instabilities, and in any case, they are
beyond the scope of our investigation.

We would like to emphasize once more that though we used a small value of $p$ in our examples, the effect of higher order derivatives in the equation of motion is still present and does play an important role. Let us make this more clear. If we had started from action (\ref{cutaction1}) that only considers the first order term in $\Box$, and expanding using $\psi = 1+\phi$, we would have obtained the equation of motion
\be
\label{example1}
\Box\phi = -\frac{2}{\alpha} \phi \,.
\ee
However, the local equation of motion (\ref{LocalField}) that follows from the full linearised non-local  equation of motion
(\ref{eqm1}) gives instead
\be
\Box \phi = - \frac{\ln p}{\alpha} \phi \,,
\ee
which, though similar for small $p$, is clearly different from Eq.~(\ref{example1}) and leads to great disparity in the results for large $p$.

\subsection{Cubic superstring field theory (CSSFT)}
\begin{figure}
\includegraphics[width=8.5cm]{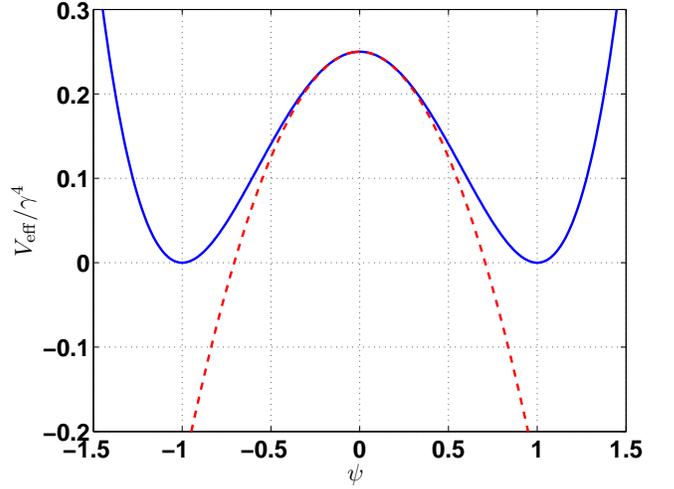}
\caption{\label{veff2} The form of the effective potential for $\psi$ in the CSSFT case. The solid line and dashed lines represent the full potential and the quadratically expanded potential around the hilltop, respectively.}
\end{figure}

In the CSSFT case we have $\lambda = -1$, $\xi^2 = 0.95$, $p = 3$, and
\begin{eqnarray}
\alpha = \frac{1}{4 m_s^2} \,, \hspace{1cm} 
\gamma^4 = \frac{m_s^4}{g_s^2}  \,.
\end{eqnarray}
Expanding the action (\ref{Lag}) to first order in $\Box$ , we obtain 
\be
S_\psi = \int d^4 x \sqrt{-g} \left( \frac{1}{2}\gamma^4 \alpha (4\xi^2-1) \psi \Box \psi - 
V_{\rm eff}(\psi) \right) \,.
\ee
Hence, for $\xi^2 > 1/4$, similarly to what happens in the $p$-adic case, we also have a non-phantom field with non-trivial kinetic terms.
The effective potential is in this case
$V_{\rm eff}(\psi) = -\gamma^4(V(\psi) + \psi^2/2)$, where $\Lambda$ is a constant related to the D-brane tension and we set its value such that $V_{\rm eff}(1) = 0$. This potential is shown in  Fig.~\ref{veff2}. 

The numerical procedure to solve the diffusion equation in this model follows closely the method described for the $p$-adic string, but since now the hilltop is at $\psi = 0$ we set $A = \epsilon$ and hence, we specify the initial data as
\begin{eqnarray}
\label{profile3}
\Psi(t_{\rm i},r) &=&  \epsilon \, e^{\alpha\omega^2_R \, r} 
\cos(\alpha\omega^2_I r) \,, \\
\label{profile4}
\dot{\Psi}(t_{\rm i},r) &=& 0 \,,
\end{eqnarray}
with $\epsilon=10^{-3}$, $\omega^2$ is a root given by Eq.~(\ref{aw2c}),
$\alpha = 1/(36 \times 10^{-12} \,\xi^2)$ and $\gamma^4 = 1.48\times10^{-9}$ in Planck units.
For $\xi^2 \neq 0$, however,
the imaginary parts of $\omega^2$ do vary with the expansion point $A$ but only by a 
significant  amount when $n=0$ or $n= -1$. In these cases, the imaginary parts only exist for 
$|A| \gtrsim 0.7$ and correspond to small frequencies as $\alpha \omega^2_I < 1$, hence, one period is not completed between $0<r<1$. For $n \geq 1$ and $n < -1$, the frequency of the profile is nearly independent of $A$. In this model, therefore, we also use as a fitting smoothing function the profile (\ref{psifit}).

\begin{figure}
\includegraphics[width=8.5cm]{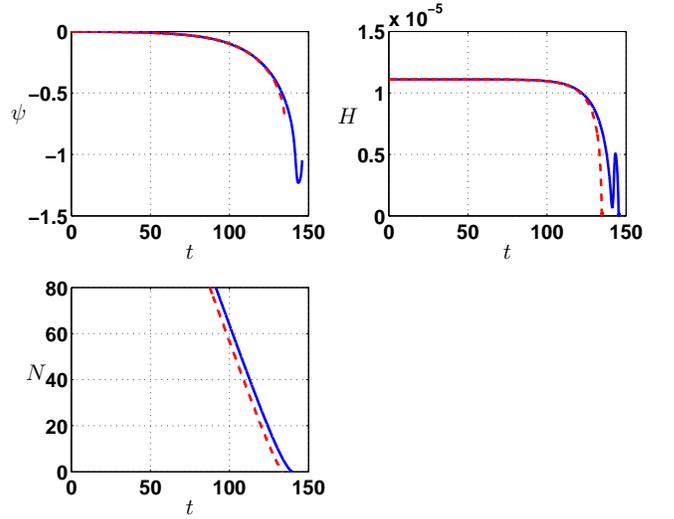}
\caption{\label{cssftres1} Upper left panel: Evolution of the field $\psi$ in the CSSFT case, $\xi^2=0.95$, $p=3$ where $\psi$ is originally a single real field, $n=0$, but develops a quintom for 
$|\psi| \gtrsim 0.7$. The solid and dashed lines represent the non-linear and linear evolution, respectively. Upper right panel: The corresponding evolution of the Hubble rate. Lower left panel: The number of $e$-folds before the end of inflation. }
\end{figure}

The results of integrating the diffusion equation numerically are shown in 
Fig.~\ref{cssftres1}. Starting the field near the hilltop corresponds,
as we have seen, to a real field. Once it has been displaced beyond
$|\psi| \approx 0.7$ it develops a quintom as $\alpha\omega^2$ becomes
imaginary. We indeed see that the Hubble rate performs an oscillation
when the field is near $|\psi| = 1$ and eventually collapses to 
$H = 0$. As in the $p$-adic case, the non-linear analysis interpolates between the linear expansion behavior about the hilltop and the 
linear expansion evolution about the minimum, where in this case, only quintoms are found. 

\begin{figure}[]
\includegraphics[width=8.5cm]{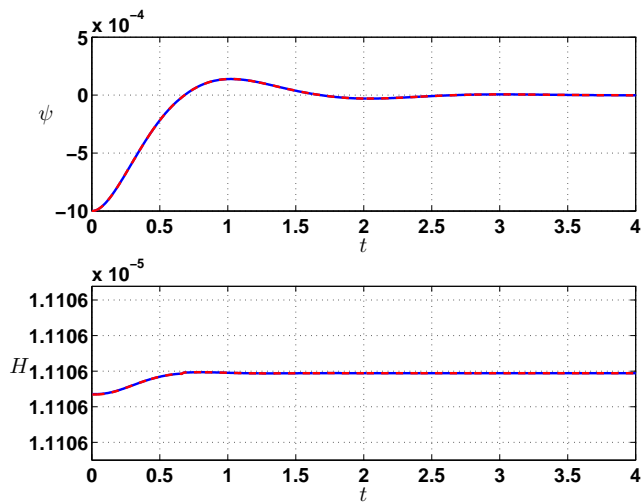}
\caption{\label{cssftres2} Upper panel: Evolution of the field $\psi$ in the CSSFT case, $\xi^2=0.95$, $p=3$ where $\psi$ is a quintom with $n=1$. The solid and dashed lines represent the linear and non-linear evolution, respectively. Lower panel: The corresponding evolution of the Hubble rate.}
\end{figure}

As we have seen in Section \ref{lineareqns}, when $m^2 = 0$ exactly, there are no quintoms allowed as the only root is real. But slightly away from the hilltop, however, there can exist quintoms as the root can be complex (see Eq.~(\ref{aw2c})). 
The evolution for the quintom with $n=1$, is shown in Fig.~\ref{cssftres2}. Its behavior is somewhat surprising given that it is fairly distinct from the evolution of the quintoms in the $p$-adic string (in Fig.~\ref{padicres2}). Instead of oscillating around the hilltop with growing amplitude, the amplitude of the field decays around the hilltop. This can be explained by noticing that the real part of $\omega^2$ is negative and becomes larger in absolute value when the field is near the hilltop whereas its imaginary part is small and nearly constant. This implies that  
both modes in solution (\ref{DiffSolveGen}) are decaying.
In this example, the field gets trapped  at $\psi = 0$ and consequently the expansion of the universe becomes closer to de Sitter. Though this quintom is a viable way of obtaining accelerated expansion, an additional mechanism to exit inflation and enter the radiation epoch is required to complete the scenario.

\section{An aside on perturbations}
\label{Perts}
Our analysis of non-local dynamics can readily be extended to perturbations about 
the homogeneous background evolution.  We leave a full analysis to future work, and simply make some observations here.

First, we note that for linear non-local theories 
the localization technique works equally well when a perturbation is 
added to the background field and to the metric. This is valid because the technique 
uses an unspecified  $\Box$ operator.  Then the perturbed linear non-local case 
is simply equivalent to a perturbed infinite field local cosmology. 
The diffusion-like equation approach to the perturbed linear non-local theory must therefore also be tractable in this case since it is equivalent to the localization technique. For the non-linear 
case the situation is more involved but a similar idea to that
employed for the background is possible. That is, we can use the
initial profiles determined in the linear case as approximate initial
conditions for the non-linear case.

As a first approximation one might consider perturbing the non-local 
scalar field and neglecting the inevitable back-reaction on the metric keeping it unperturbed. Separating out the unperturbed equations, and 
decomposing the field perturbation into Fourier modes, 
the non-local scalar field equation for a particular mode takes the form 
\be
\label{Eqpert}
F(\Box_k)\delta\psi_k=\frac{\partial^2 V}{\partial \psi^2} \delta \psi_k \,~,
\ee
where $k$ labels the wave number of the mode, and where $\Box_k$ is 
\be
\Box_k =-\frac{d^2 }{ dt^2}-3 H \frac{d}{dt}- \frac{k^2}{a^2} \,.
\ee
$H$ and $a$ are simply the Hubble rate and scale factor of the unperturbed universe respectively.

The first thing to note is that this equation is linear, but the situation is not as straightforward as in the linear unperturbed case as the effective mass 
($\partial^2 V/\partial \psi^2)$ is now time dependent. This means that the correspondence technique 
does not work.  We could approximate the effective mass as a constant for a short period 
of evolution, for example, as a mode exits the cosmological horizon but this 
procedure is bond to be inaccurate. 
An alternative is to use the diffusion-like equation technique. We can 
introduce the new field 
\be
\delta \Psi(t,r) = e^{-r \alpha\Box_k} \delta\psi_k(t) \,, 
\ee
which satisfies the equation
\be
\Box_k \delta\Psi =- \frac{1}{\alpha} \, \frac{\partial \,\delta\Psi}{\partial r} \,.
\ee
For the potential $V(\psi) = - \psi^4/4$ and $\xi^2 = 0$, 
the field equation for the perturbation translates 
into the boundary condition
\be
\label{pertBoundaryLinear}
\delta \Psi (t,1)= 3 \Psi(t,0)^2 \, \delta\Psi(t,0) \,,
\ee
which is linear but time dependent.  
As for the unperturbed field, we must specify 
$\Psi(t_{\rm i},r)$ and $\dot{\Psi}(t_{\rm i},r)$
which must also satisfy the boundary conditions (\ref{BoundaryMany1}), if we are to solve the 
diffusion equation as a initial value problem.  However, a 
very similar problem arises as in the 
non-linear background case.  That is, our single boundary 
condition in the $r$ direction 
actually imposes an infinite set of boundary condition. Indeed, the second boundary condition 
with $\dot{\Psi}(t_{\rm i},r) =\delta \dot{\Psi}(t_{\rm i},r) = 0$
reads
\begin{eqnarray}
\left[\frac{\partial \, \delta\Psi(t_{\rm i},r)}{\partial r}\right]_{r = 1} &=&
6 \Psi(t_{\rm i},0) \left[\frac{\partial \, \Psi(t_{\rm i},r)}{\partial r}\right]_{r = 0} \delta \Psi(t_{\rm i},0) \nonumber \\
&~& + 3 \Psi(t_{\rm i},r)^2 \left[\frac{\partial \, \delta \Psi(t_{\rm i},r)}{\partial r}\right]_{r = 0} \,.
\end{eqnarray}
We can only satisfy these approximately as we did for the non-linear
unperturbed case, and then use the diffusion-like equation to evolve
forward in time.  While we leave such an analysis to future work, we mention
it here to highlight the utility of the initial value formulation of
the diffusion equation method when compared with the boundary value
formulation. Of course one could also obtain a solution using the boundary
value formulation, but for that procedure 
$\delta\Psi$ must tend to a constant at $\pm \infty$ fixed by 
Eq.~(\ref{Eqpert}). 
For example, for the $p$-adic case we must have  
$\delta \psi_k =3 \psi^2 \delta \psi_k$ at $t = \pm \infty$ as all the derivatives vanish there.  Since $\psi$ can only take values 
$\psi = \pm 1$,0 at $t= \pm \infty$, it is clear that $\delta \psi_k=0$. 
Hence the boundary value formulation 
can only generate solutions in which the perturbation
decays 
asymptotically. However, in a study of perturbations produced
during non-local inflation for example, we would like to put a particular
$k$-mode of the  perturbation
into a state similar to the Bunch-Davis vacuum initially and evolve it
forward in time until it exits the horizon.  Repeating this for many modes we
can determine the spectrum of perturbations produced.  This utility
is a major motivation for
studying the diffusion-like pde technique from the point of view of an
initial value problem.

\section{Conclusions and future directions}
In this work, our major concern has been whether non-local cosmologies
can be solved as an initial value problem by solving a related
partial differential equation as an initial value problem.  We have
found that this is indeed possible but that there are technical
difficulties, both in understanding how to specify suitable initial data
and in overcoming the ill-posedness of the pde. 

The first difficultly is easily resolved in linear models where
separation of variables leads to an interpretation of the non-local
cosmology in terms of a local one with an infinite number of fields.
We showed, intriguingly, that most of these fields are in the form of
quintoms (one standard field and a coupled phantom field).  Moreover, the
separation of variables technique led to a simple understanding of
suitable initial data for the diffusion-like equation.  In non-linear cases, we showed that
linearising the equations allowed approximate initial data to be
specified for a subsequent non-linear evolution.  Extending previous
analysis, we showed that this linearisation can be preformed about regions
of the potential other than a maximum or minimum.

The second difficultly we overcame by using a simple regularization
technique, where at each time step we smoothed the profile of the
solution in the $r$ direction to eliminate spurious oscillations which
would otherwise swamp the solution.  The smoothing which worked best
took advantage of the theoretical knowledge about the probable form of the
profile (from the linearised analysis), though we stress that the
free parameters in the profile were fitted from the data rather than imposed. 

Previous work on inflation in non-local theories has been carried out by a
number of authors. A major conclusion of our work is that the method
employed in Ref.~\cite{lidsey} (i.e. the linearisation  of the equations about the
hill-top, considering the evolution of only the real canonically
normalized local field) works very well for the
background evolution when observationally  
relevant scales are exiting the cosmological horizon in the cases we have studied. Though we note that the evolution can be very 
different if other local fields (the quintoms) are permitted to be
present initially. Such a localization method allows known results for
a canonically normalized field to be employed, and hence the power
spectrum can be calculated for the non-local theory.  Of course,
before making a firm statement we would have to perform the analysis 
presented in
this article for perturbations as well as for the background evolution. 
We sketched how this could be achieved in Section \ref{Perts}  and we will return to
this issue in future work.  Many questions should be rigorously addressed, for
example: In a non-local theory, is the comoving curvature perturbation
still conserved on super-horizon scales even when the evolution is
non-linear? This needs to be known before the previous results (and in particular claims of a 
large non-gaussianity in $p$-adic inflation when $p$ is large \cite{non-Gaussian}) can be
confirmed or falsified as they all relay on it. 

Since several authors have also considered non-local tachyons as the
source of dark energy in the universe \cite{darkTachyon}, another obvious
application of our methodology is to include a background fluid into
these models and see how the field behaves during radiation and mater
domination.  Only when this is known can a firm statement on whether a
non-local field is a dark energy candidate can be made.

\begin{acknowledgments}
D.J.M is supported by the Centre for Theoretical Cosmology and N.J.N by STFC. The authors thank James Lidsey, Neil Turok, Daniel Wesley and David Seery for discussions and Liudmila Joukovskaya for numerous conversations and comments on the manuscript. 
\end{acknowledgments}


\begin{thebibliography}{99}


\bibitem{NonLocalCosmolOthers}

  J.~Khoury,
  Phys.\ Rev.\  D {\bf 76}, 123513 (2007)
  [arXiv:hep-th/0612052];
  
  N.~Barnaby, T.~Biswas and J.~M.~Cline,
  JHEP {\bf 0704}, 056 (2007)
  [arXiv:hep-th/0612230];
 
 I.~Y.~Aref'eva and I.~V.~Volovich,
  arXiv:hep-th/0612098;
 
 I.~Y.~Aref'eva, L.~V.~Joukovskaya and S.~Y.~Vernov,
  JHEP {\bf 0707}, 087 (2007)
  [arXiv:hep-th/0701184];
  
 G.~Calcagni,
  JHEP {\bf 0605}, 012 (2006)
  [arXiv:hep-th/0512259];

 I.~Y.~Aref'eva and L.~V.~Joukovskaya,
  JHEP {\bf 0510}, 087 (2005)
  [arXiv:hep-th/0504200].


\bibitem{darkTachyon}
I.~Y.~Aref'eva,
  AIP Conf.\ Proc.\  {\bf 826}, 301 (2006)
  [arXiv:astro-ph/0410443].

 I.~Y.~Aref'eva, A.~S.~Koshelev and S.~Y.~Vernov,
  Phys.\ Rev.\  D {\bf 72}, 064017 (2005)
  [arXiv:astro-ph/0507067];

 I.~Y.~Aref'eva and A.~S.~Koshelev,
  arXiv:0804.3570 [hep-th].

\bibitem{koshelev}
A.~S.~Koshelev,
  JHEP {\bf 0704}, 029 (2007)
  [arXiv:hep-th/0701103].

\bibitem{calcagni1}
G.~Calcagni, M.~Montobbio and G.~Nardelli,
  Phys.\ Rev.\  D {\bf 76}, 126001 (2007)
  [arXiv:0705.3043 [hep-th]]; 

\bibitem{calcagni2}

  Phys.\ Lett.\  B {\bf 662}, 285 (2008)
  [arXiv:0712.2237 [hep-th]];  G.~Calcagni and G.~Nardelli,
  arXiv:0802.4395 [hep-th].


\bibitem{lidsey}
  J.~E.~Lidsey,
  Phys.\ Rev.\  D {\bf 76}, 043511 (2007)
  [arXiv:hep-th/0703007].


\bibitem{non-Gaussian}
  N.~Barnaby and J.~M.~Cline,
  JCAP {\bf 0707}, 017 (2007)
  [arXiv:0704.3426 [hep-th]];
 
  N.~Barnaby and J.~M.~Cline,
  arXiv:0802.3218 [hep-th].


\bibitem{ludaLinear}

  I.~Y.~Aref'eva, L.~V.~Joukovskaya and S.~Y.~Vernov,
  arXiv:0711.1364 [hep-th].

\bibitem{ludaBoundary}
  L.~Joukovskaya,
  Phys.\ Rev.\  D {\bf 76}, 105007 (2007) 
[arXiv:0707.1545 [hep-th]];
  arXiv:0803.3484 [hep-th];
  arXiv:0807.2065 [hep-th].

\bibitem{biswas}

  T.~Biswas, A.~Mazumdar and W.~Siegel,
  JCAP {\bf 0603}, 009 (2006)
  [arXiv:hep-th/0508194].



\bibitem{sft}
M. Kaku and K. Kikkawa,
Phys. Rev. D {\bf 10}, 1110 (1974); 
E. Witten, Nucl. Phys. {\bf B268}, 253 (1986). 

\bibitem{reviews}
K. Ohmori, hep-th/0102085; 
I. Ya. Aref'eva, D. M. Belov, A. A. Giryavets, A. S. Koshelev, and 
P. B. Medvedev, hep-th/0111208; 
A. Sen, Int. J. Mod. Phys. {\bf A20}, 5513 (2005), hep-th/0410103.  

\bibitem{p-adic}
P. G. O. Freund and M. Olson, 
Phys. Lett. {\bf B199}, 186 (1987); 
P. G. O. Freund and E. Witten, Phys. Lett. {\bf B199}, 191 (1987); 
L. Brekke, P. G. O. Freund, M. Olsen, and E. Witten, 
Nucl. Phys. {\bf B302}, 365 (1988). 

\bibitem{vladimirov}
V.~S. Vladimirov, arXiv:math-ph/0507018.

\bibitem{ivp}
  N.~Barnaby and N.~Kamran,
  JHEP {\bf 0802}, 008 (2008)
  [arXiv:0709.3968 [hep-th]].

\bibitem{se}
H.~t.~Yang,
  JHEP {\bf 0211}, 007 (2002)
  [arXiv:hep-th/0209197].



\bibitem{Caldwell:1999ew}
  R.~R.~Caldwell,
  Phys.\ Lett.\  B {\bf 545}, 23 (2002)
  [arXiv:astro-ph/9908168].

\bibitem{quintoms}
  B.~Feng, X.~L.~Wang and X.~M.~Zhang,
  Phys.\ Lett.\  B {\bf 607}, 35 (2005)
  [arXiv:astro-ph/0404224].

\bibitem{Carroll:2003st}
  S.~M.~Carroll, M.~Hoffman and M.~Trodden,
  Phys.\ Rev.\  D {\bf 68}, 023509 (2003)
  [arXiv:astro-ph/0301273].

\bibitem{ostro}
M. Ostrogradski, Mem. Ac. St. Petersbourgh VI, {\bf 4}, 385 (1850).

\bibitem{moeller}
  N.~Moeller and B.~Zwiebach,
  JHEP {\bf 0210}, 034 (2002)
  [arXiv:hep-th/0207107].

\bibitem{volovich} 
 Y.~Volovich,
  J.\ Phys.\ A  {\bf 36}, 8685 (2003)
  [arXiv:math-ph/0301028].

\bibitem{Hellerman:2008wp}
  S.~Hellerman and M.~Schnabl,
  arXiv:0803.1184 [hep-th].

\end{thebibliography}
\end{document}